# Multidimensional mutual information methods for the analysis of covariation in multiple sequence alignments

**Greg W. Clark [1*], Sharon H. Ackerman [2*], Elisabeth R. Tillier [1§], and Domenico L. Gatti [2,3, §]**

[1]Department of Medical Biophysics, University of Toronto, Campbell Family Institute for Cancer Research, Ontario Cancer Institute, University Health Network, Toronto, Ontario, Canada.
[2]Department of Biochemistry and Molecular Biology, Wayne State University School of Medicine, Detroit, Michigan, USA
[3]Cardiovascular Research Institute, Wayne State University School of Medicine, Detroit, Michigan, USA

[§]Corresponding author
[*]These authors contributed equally to the study.

Email addresses:
    GWC: gregw.clark@mail.utoronto.ca
    SHA: sackerm@med.wayne.edu
    ERT: e.tillier@utoronto.ca
    DLG: dgatti@med.wayne.edu

## Abstract

  Several methods are available for the detection of covarying positions from a multiple sequence alignment (MSA). If the MSA contains a large number of sequences, information about the proximities between residues derived from covariation maps can be sufficient to predict a protein fold. However, in many cases the structure is already known, and information on the covarying positions can be valuable to understand the protein mechanism and dynamic properties.

  In this study we have sought to determine whether a multivariate (multidimensional) extension of traditional mutual information (MI) can be an additional tool to study covariation. The performance of two multidimensional MI (mdMI) methods, designed to remove the effect of ternary/quaternary interdependencies, was tested with a set of 9 MSAs each containing <400 sequences, and was shown to be comparable to that of the newest methods based on maximum entropy/pseudolikelyhood statistical models of protein sequences. However, while all the methods tested detected a similar number of covarying pairs among the residues separated by < 8 Å in the reference X-ray structures, there was on average less than 65% overlap between the top scoring pairs detected by methods that are based on different principles.

  We have also attempted to identify whether the difference in performance among covariation detection methods is due to different efficiency in removing covariation originating from chains of structural contacts. We found that the reason why methods that derive partial correlation between the columns of a MSA provide a better recognition of close contacts is not because they remove chaining effects, but because they filter out the correlation between distant residues that originates from general fitness constraints. In contrast we found that true chaining effects are expression of real physical perturbations that propagate inside proteins, and therefore are not removed by the derivation of partial correlation between variables.

## Background

During the past ten years there has been significant progress in the development of computational tools for the detection of co-evolution between pairs of positions in a protein family by analysis of its MSA (reviewed in [1-5]). If the MSA of a protein family contains a sufficiently large number of sequences, information about the proximities between residues derived from the covariation map can be used to predict the protein fold. However, in many cases the structure of one or more members of a protein family is already known, and interest in identifying covarying positions lies instead in the information that this knowledge can provide about the protein mechanism and dynamic properties, or in its use as a starting point for mutagenesis studies.

Unfortunately, the reliability of covariation data can be diminished by the existence of correlations originating not just from the direct interactions (physical or functional) between two residues, but also from their shared interaction with one or more other residues, and by the shared phylogenetic history of several homologous proteins in the MSA. Attempts to disentangle direct from indirect and phylogenetic correlations were made with the MIp/APC [6], Zres [7] and Zpx [8] corrections of MI statistics, with the application of Bayesian modeling in the logR method [9], with Direct Coupling Analysis (DCA) [10-13], a maximum entropy method, with the use of sparse inverse covariance estimation in the PSICOV method [14, 15], and most recently using a pseudolikelyhood framework [16-18], or combining principal component analysis (PCA) with DCA [19]. While the performance of these methods has been tested primarily with high quality MSAs containing a very large number of sequences (between 5L and 25L, with L = sequence length), very often investigators are interested in studying the covarying positions of proteins for which the available MSA contains less than L sequences, and whose alignment quality is not optimal due to the presence of many (or large) gaps, or significant sequence heterogeneity in the protein family. In these cases, it is difficult to argue that a single best method exists, since different algorithms may be more (or less) effective in capturing the covariation signal from MSAs with widely different statistical properties, and a better strategy may rely on merging the information derived from a few methods based on different principles. In order to expand the choice of algorithms available for covariation analysis, here we present a new class of methods, based on multidimensional mutual information (mdMI), specifically designed to remove indirect coupling up to ternary/quaternary interdependencies. These new methods were tested on a set of 9 protein families each represented by a MSA containing between ~0.4 and ~2L sequences.

## Results

### Derivation of 3D and 4D MI covariation matrices

In most traditional applications mutual information is used to study the interaction between two variables. If we consider a channel with a single discrete input X1 and a single discrete output X2, the amount of transmission between X1 and X2 is defined as their 'mutual information' I(X1;X2):

$$I(X1;X2) = H(X1) + H(X2) - H(X1,X2) \tag{1}$$

where the H's represent the individual (X1 or X2) and joint (X1,X2) entropies. For our application, X1 and X2 represent columns in the MSA. We can consider a more complicated case including a third channel (column). In this case, I(X1;X3;X2) between the three variables represents the 'interaction information' for a channel with two discrete inputs X1 and X3 and a single discrete output X2 (a 2-way channel). It is defined [20] as :

$$I(X1; X2; X3) = {\color{red}H(X1) + H(X2) + H(X3) - H(X1, X2)} \\ {\color{blue}- H(X1, X3) - H(X2, X3) + H(X1, X2, X3)} \tag{2}$$

The mutual information I(X1;X2) is provided by the red terms in (2). The following considerations help understand the nature of the remaining terms. If we are interested in 'explaining out' the effect of X3 on the transmission between X1 and X2, we can take a sum of the mutual information I(X1;X2) for each possible value x3 of X3, weighted by the probability of occurrence ($p_{x3}$) of each of those values:

$$I_{X3}(X1; X2) = \sum p_{x3}\, I(X1; X2 \mid X3 = x3) = I(X1; X2 \mid X3)$$
$$= I(X1, X3 ; X2) - I(X3 ; X2)$$

where I(X1, X3 ; X2) is the MI between the joint variable (X1,X3) and X2. Developing further we have:

$$= [H(X1, X3) + H(X2) - H(X1, X3, X2)] - [H(X3) + H(X2) - H(X3, X2)]$$
$$= H(X1, X3) - H(X1, X3, X2) - H(X3) + H(X3, X2) \qquad (3)$$

the right-hand side of (3) can be recognized as the negative of the blue terms in (2). From this observation and rearranging, we obtain that the mutual information $I_{X3}$(X1; X2) between X1 and X2, when the effect of X3 on the transmission between them has been eliminated, can be obtained by subtracting the interaction information I(X1; X2; X3) from the mutual information I(X1;X2):

$$I_{X3}(X1; X2) = I(X1; X2) - I(X1; X2; X3) \qquad (4)$$

Averaging over all values of X3 (a 3rd column) in an MSA we obtain for the 3-dimensional MI between any two columns (X1 and X2):
$$<I_{X3}(X1; X2)>_{X3 \neq X1, X2} = I(X1; X2) - <I(X1; X2; X3)>_{X3 \neq X1, X2}$$
$$= <H(X1, X3) - H(X1, X3, X2) - H(X3) + H(X3, X2)>_{X3 \neq X1, X2} \qquad (5)$$

Since the effects of all such 3rd residues are averaged, 3-dimensional MI provides a *global* removal of all indirect couplings exerted on a pair by any other individual residue in the sequence (ternary interdependencies).

Likewise, the mutual information $I_{X3,X4}$(X1; X2) between X1 and X2, when the effect of two additional variables X3 and X4 on the transmission between them is removed, is obtained [21, 22] as:

$$I_{X3,X4}(X1; X2) = \sum p_{x3,x4}\, I(X1; X2 \mid X3 = x3, X4 = x4)$$
$$= I(X1; X2 \mid X3, X4) \qquad (6)$$

By the 'chain property' of multivariate MI [23] we derive:

$$I(X1; X2 \mid X3, X4) = I(X1; X2 \mid X4) - I(X1; X2; X3 \mid X4)$$
$$= I(X1,X2) - I(X1;X2;X4) - I(X1;X2;X3) + I(X1;X2;X3;X4) \qquad (7)$$

where the 'interaction information' between the four variables is:

$$I(X1;X3;X2;X4) = [H(X1) + H(X2) + H(X3) + H(X4)] - [H(X1,X2) + H(X1,X3) + H(X1,X4) + H(X2, X3) + H(X2,X4) + H(X3,X4)] + [H(X1,X2,X3) + H(X1,X2,X4) + H(X1,X3,X4) + H(X2,X3,X4)] - H(X1,X2,X3,X4) \qquad (8)$$

Averaging over all values of X3 and X4 (two additional columns of the MSA), and recalling that all the values taken by X3 and X4 are the same with respect to X1 and X2 in a MSA (although X3 $\neq$ X4), we finally obtain:

$$\langle I_{X3,X4}(X1;X2)\rangle_{X3,X4\neq X1,X2} = I(X1;X2) - 2\langle I(X1;X2;X3)\rangle_{X3\neq X1,X2} +$$
$$\langle I(X1;X2;X3;X4)\rangle_{X3,X4\neq X1,X2} \qquad (9)$$

Expanding (7) leads to a direct expression of $I_{X3,X4}(X1;X2)$ in terms of the entropies of the individual variables and simplifies to (**Text S1**, Supporting Information):

$$I_{X3,X4}(X1;X2) = -H(X3,X4) + H(X1,X3,X4) + H(X2,X3,X4) - H(X1,X2,X3,X4) \qquad (10)$$

Averaging over X3 and X4:

$$\langle I_{X3,X4}(X1;X2)\rangle_{X3,X4\neq X1,X2} = \langle -H(X3,X4) + H(X1,X3,X4)$$
$$+ H(X2,X3,X4) - H(X1,X2,X3,X4)\rangle_{X3,X4\neq X1,X2} \qquad (11)$$

Since the effect of any 3[rd] and 4[th] residues are averaged, 4-dimensional MI provides a *global* removal of all indirect couplings exerted on a pair by any two other residues in the sequence (quaternary interdependencies).

Both (5) and (11) can be computed from the marginal frequencies of the aa symbols in any 3 or 4 columns of a MSA.

We have implemented equations (5,11) as a Matlab function (NMSA_to_mdMI) included in the new release of our Toolbox (MSAvolve_v2.0a, which can be downloaded from http://146.9.23.191/~gatti/coevolution/) for the covariation analysis of MSAs. Upon derivation of the $\langle I_{X3}(X1;X2)\rangle$ and $\langle I_{X3,X4}(X1;X2)\rangle$ values, the covariation matrices are further processed as described in our earlier work [4] to derive the corresponding ZPX2 matrices [8]. The final matrices are named '3D_MI' and '4D_MI' (for the 3- and 4- dimensional cases, respectively). The same function provides also a standard MI map, which for consistency is called here '2D_MI'.

In its current parallel version 3D_MI runs on a single CPU in ~3 min with a MSA of 250 positions and 300 sequences, and its speed increases almost linearly with the recruitment of more cpu's. As expected 4D_ZPX2 is significantly slower and with large memory requirements. Despite this limitation, 4D_MI can be very useful to analyze in great detail small parts of a large MSA. However, in the following section we show that the simpler and faster 3D_MI is already very effective in calculating the direct coupling between the positions of a MSA.

**Prediction of close contacts in X-ray crystal structures**

We have evaluated the performance of standard MI (2D_MI), 3D_MI, 4D_MI, PSICOV [14], plmDCA [17], GREMLIN [18], and Hopfield-Potts_DCA with Principal Component Analysis [19](called here hpPCA) with the MSAs of 9 protein families, which we have used as test sets in our recent survey of covariation detection methods [4]. These MSAs contain less than 400 sequences with ratios of sequence number to sequence length (called here the 'L ratio') between 0.4 and 2.0, and thus represent a particularly sensitive test for the performance of the different methods with less than optimal size MSAs. It is worth noting that PSICOV, plmDCA, GREMLIN, and hpPCA apply by default the 'average product correction' (APC or MIp correction) that was originally introduced by Gloor [6] as a correction for the entropic and phylogenetic bias of MI statistics. Later on, independently, Little [7] and Gloor [8] introduced a second correction called respectively ZRES or ZPX, to be applied after the APC correction, and Dickson [24, 25] showed that this second correction further improves the accuracy of covariation detection particularly in MSAs containing some degree of misalignment. For this reason, an APC correction (if not already present) and a ZPX correction were applied to all covariation maps derived in this study with different methods. We found that the maps so corrected, performed uniformly better than the uncorrected maps in the detection of close contacts. Although each MSA contained less than 400 sequences, all the methods tested produced covariation maps that closely resembled the contact maps derived from the representative X-ray structures of each family: and example of these maps is shown in **Figure 1** for the MDH family, which contains 391 sequences with an L ratio slightly larger than 1.

To quantify the detection of close contacts, we measured what percentage of all residue pairs separated by less than 8 Å in the X-ray structure was represented in the top covarying pairs identified by each method. A number of pairs equal to the number of residues L in each sequence was considered. This result was further filtered to include either all the pairs or only pairs whose component residues are separated by at least 6, 12, 20 positions in sequence space. Results obtained with all 9 MSAs for all pairs and for pairs separated by at least 20 positions in sequence are shown in **Figure 2** (averaged results are shown in **Figure S1**, Supporting Information). When all possible pairs are considered (left panels of each protein family in **Figure 2**), plmDCA identifies a higher percentage of pairs separated by < 8 Å in almost all the MSAs. However, a significant variability in performance between the different methods becomes apparent when only pairs separated by at least 20 positions in sequence (right panels of each protein family) are considered. For example, in the ArsA family plmDCA clearly includes the highest percentage (~6.5%) of proximal pairs (in space) in the top 583 covariation scores (**Figure 2**, ArsA family, left panel): however only 300 out of these 583 pairs are in the subset of pairs that are distant in sequence, and these 300 represent only <0.5% of all pairs close in space (**Figure 2**, ArsA family, right panel). Conversely, GREMLIN includes a smaller number (~4.4%) of proximal pairs in the top 583 covariation scores, but 400 of them are in the subset of pairs that are distant in sequence, and they represent almost 1% of all pairs close in space. These results indicate that the better overall performance of plmDCA with ArsA (when all pairs are considered) is due to the fact that a very large number of pairs close in sequence is represented in the top 583 covariation scores. Extending this type of analysis to all 9 protein families, it becomes clear that if we were primarily interested in the pairs that are distant in sequence but close in space, we would perhaps achieve the highest accuracy using GREMLIN with ArsA, ArsC, and MDH, but 4D_MI would be the method of choice with Atp11p, PHBH, and CcrA, while either GREMLIN, plmDCA, and hpPCA would work best for Atp12p.

However, even for cases in which several methods give comparable results in the type of analysis shown in **Figure 2**, we may wonder whether the pairs identified by any two of these methods are the same or not. In **Table 1** the average percentage (including all 9 MSAs) of pairs shared by all methods in the top L covariation scores is shown as a matrix, with values above the diagonal referring to all possible pairs, and values below the diagonal referring to pairs separated by at least 20 positions in sequence. This matrix shows that a significant number of pairs are shared among methods that are conceptually similar (for example 2D_MI, 3D_MI and 4D_MI). However, the percentage of pairs shared by methods that are conceptually different is much smaller: for example 3D_MI and plmDCA share less than 40% of all pairs, and even plmDCA and GREMLIN, which operate within a similar pseudolikelyhood framework, share at most 64% of all pairs.

Thus, even from just this set of only 9 protein families, it appears that a single method that would give the best result with all protein families is hard to find, as each algorithm performs quite differently with different MSAs, and even algorithms whose overall performance is similar on a statistical basis, share no more than 2/3 of all the pairs among the top L covariation scores, if they are based on different principles.

**Dependence of the covariation signal on secondary structure**

If there are significant differences in the performance of each methods with different MSAs, it is important to understand the origin of such differences, and how they affect the structural information derived from covariation maps. For example, since each protein family is characterized by a variable mixture of secondary structures (e.g., amount, size, and orientation of helices and strands), we have analyzed the dependence of the covariation signal detected by each methods on the type(s) of secondary structure (helices, strands) in which covarying pairs are located (see Methods). Non-independent selection of neighboring residues is a phenomenon known to occur in helices and strands, and thus the features of these secondary structures provide a rich framework to study residue coupling. With the set of 9 protein families, all methods produce noticeable peaks or shoulders at periods corresponding to up to 4 helix turns, and at periods of 2, 4, and 6 residues in strands, corresponding to side-by-side residues pointing in the same direction (**Figure 3**). However each method has its own pattern with stronger covariation scores assigned on average on one or another of these periods. Thus, the wide performance range of each method with different MSAs results in contact predictions that may reflect more or less well the secondary structure features of each protein family.

**Network connectivity**

An important aspect of recent improvements in the accuracy of covariation detection methods is the separation of the direct coupling between two residues from the indirect coupling. Given residues A,B,C, when pair AB and pair BC are among the top pairs and represent true structural contacts based on protein geometry, we may find pair AC as highly covarying (yet distant in geometric structure) as an induced coupling produced by pairs AB and BC. This kind of induced coupling can extend along chains of contacts: for example if A contacts B, B contacts C, C contacts D, ...N-1 contacts N, covariation maps may show some level of covariation between A and N. This type of covariation is not necessarily a statistical artifact. In fact since proteins are very tightly packed a mutation that produces a change in volume in some part of the structure, can be compensated by small changes of volume along chains of residues contacting each other. A similar effect can be observed for a mutation that produces a change of charge, polarity, or hydrogen bond patterns. The very existence of this type of chaining effects as a real physical phenomenon occurring inside proteins is proven by the dependence of the covariation signal on secondary structure, which was analyzed in the previous section. True chain-dependent coupling between residues should be distinguished from the statistical correlation that may occurr between residue A and residue B because of their physical correlation to a third residue C despite the lack of any physical correlation between A and B.

We have used tools from graph theory as applied to the analysis of networks to explore further the influence of chaining effects on the covariation scores of pairs that are not in direct physical contact. We recall here that a covariation map is a weigthed adjacency matrix of the graph representing the network of interactions between residues in a protein. For each pair A,B in the covariation map, within a given threshold of top covariation scores, we ask the question: is there a physical path that connects A to B through residues that belong to high scoring pairs? If there is such a path (there could be more than one) we want to know what is the total length of the shortest path (excluding a possible direct contact between the two residues) and what is the mean length of the steps that lead from A to B. In practice this question is answered by selecting a group of pairs based on a score threshold from the covariation map of a protein family, and solving the 'traveling salesman' problem for the corresponding pairs in the distance map of a reference X-ray structure. For example, if the shortest non-direct path 20-91-123-203-78 is found between the components of the high scoring pair 20-78, it means that pairs 20-91, 91-123, 123-203, and 203-78 are present among the pairs selected within a given number of top covariation scores. An example of this type of analysis is presented in **Figure 4A** for the MDH protein family using covariation maps derived with different methods. The top right panel shows the mean length of the path connecting the pairs within scoring threshold: the threshold is progressively lowered to include a larger number of pairs up to 3L. As more pairs are included in the analysis the probability of finding a shorter path increases and all traces converge to a smaller path length. The top left panel shows the mean length of the steps in each path. Clearly, smaller steps favor the transfer of physical perturbations (of size, charge, or other property) along a chain of contacts. It is significant that the ranking of different methods in this panel, based on the length of the steps (shorter steps ≈ more distance induced correlation), corresponds quite well to their capacity to recognize close contacts in the reference X-ray structure of the MDH family as shown in **Figure 2**. This correlation was found to hold true also for the other MSAs analyzed, and suggests that chaining effects actually favor, rather than confound, the recognition of close contacts by covariation methods. Further support for this conclusion is provided in the lower left panel of **Figure 4A**, which shows the correlation between the covariation score of pairs and the mean step length of the path that connect the two members of each pairs through residues that belong to other pairs. With the exception of a few points representing the very top scoring pairs (~L/5), most points of the best performing methods (based on **Figure 2**) sit in a range of negative correlation, indicating that the covariation score is higher when the mean step length (or also the mean path length, right panel) is smaller.

An average of the results obtained with the path length connectivity analysis for all 9 protein families is shown in **Figure S2** (Supporting Information). Also in this case, the ranking of different methods based on the mean step length of the connectivity paths (top left panel) corresponds reasonably well to the average performance of the methods in predicting close contacts (**Figure S1**, Supporting Information).

In a different type of analysis, we used the concept of graph transitivity to investigate how the covariation scores produced by different methods are affected by the statistical correlation that may occur between residue A and residue B because of their physical correlation to a third residue C despite the lack of any physical correlation between A and B. At each given threshold of covariation scores, Transitivity, *Tr*, measures the number of "completed triads" relative to the number of "potential triads", and ranges from 0 to 1 (completely connected graph). Protein covariation graphs were manipulated for the set of 9 proteins either by changing the covariation stringency (number of top pairs included in the graph based on a covariation score cutoff) or by changing the minimum sequence distance of residue pairs at constant graph size (see Methods). Transitivity was then calculated for each covariation graph.

An example of this analysis for the MDH protein family is shown in **Figure 4B**. As the graph size is increased by lowering the score threshold at a constant sequence distance of 6, transitivity values decrease for plmDCA, hpPCA, GREMLIN and PSICOV, while they increase for the three MI/mdMI based methods (2D, 3D, 4D) (**Figure 4B, left panel**). For all methods transitivity values reach a plateau when the network reaches size ≈ L (yellow vertical line). As the minimum sequence distance is increased at constant network size = L (**Figure 4B, right panel**), all methods show a trend of decreasing transitivity, with the decrement rate being maximal within the first few positions for most methods. In both types of analysis there is a clear separation of the different methods in two groups with PSICOV, plmDCA, GREMLIN and hpPCA showing significantly lower values of transitivity than the MI/mdMI based methods. The transitivity trends shown in **Figure 4B** are consistent with the path connectivity analysis shown in **Figure 4A**: top scoring pairs identified by MI/mdMI methods are connected on average by shorter indirect paths involving other residues, but the steps of these paths are longer (~13-15 Å) leading to a higher number of completed triads (as opposed to quartets, quintets, etc.). At the same time if, for example, A is connected to B and B is connected to C, the increased step length decreases the probability of a real physical perturbation propagating from A to C through B. In this case increased transitivity can be rationalized as evidence of a statistical correlation between A and C without physical interaction. However, high transitivity appears to have little effect on the identification of close contacts among residues separated by > 20 positions in sequence, as the performance of 3D_MI and 4D_MI in this respect approaches that of plmDCA or GREMLIN (**Figure 2**, MDH panel).

## Conclusions

In this study we have introduced a new class of methods to detect covariation from experimental MSAs, based on multidimensional mutual information (mdMI). Simple algebraic relationships (equations 5,11) for the removal of 3rd and 4th order indirect coupling between the columns of a MSA were derived and implemented as Matlab functions. Due to the long execution times and large memory requirements (growing with the 4th power of the sequence length) of 4D_MI only the removal of 3rd order indirect coupling (3D_MI) is practical with desktop computers for MSAs of sequences longer than 200 residues. The performance of 3D_MI and 4D_MI *vis-a-vis* that of PSICOV, plmDCA, GREMLIN and hpPCA was tested with the MSAs of 9 protein families;

although each MSA contained less than 400 sequences, all three methods produced covariation maps that closely resembled the contact maps derived from the representative X-ray structures of each family (**Figure 1**). While we observed significant variability in the performance of the methods with each MSA (**Figure 2**), on average removal of only 3rd order indirect coupling by 3D_MI was sufficient to replicate the performance of plmDCA and GREMLIN (**Figure S1**, Supporting Information). One merit of 3D_MI is its purely algebraic simplicity (see equation 5), grounded in the traditional relationships of multivariate information theory [20-22, 26].

While all the methods used in this study performed quite well in terms of percentage of close contacts recognized among the top covarying pairs, they did not necessarily recognize the same close contacts, as no more than 50% of all the pairs were shared between the MI/mdMI based methods and the other methods (**Table 1**). The percentage of pairs shared among the residues separated by at least 20 intervening positions in sequence space was even lower (~40%). Furthermore, while on average all methods produced noticeable peaks or shoulders at periods corresponding to up to 4 helix turns, and at periods of 2, 4, and 6 residues in strands, there were significant differences in the methods capacity to identify distance dependent covariation among residues located in the same secondary structures (**Figure 3**). These results suggest that if the goal of a covariation analysis is not that of structure prediction, and if one or more representative X-ray structures are available for a given protein family, analyzing both the accuracy of residue-residue contact prediction, and the patterns of distance dependent covariation in secondary structures, may point to the method(s) that offer the best performance with a specific MSA. Finally, since there is < 65% overlap among the sets of covarying residues identified by algorithms based on different principles, further improvement in accuracy is likely to be obtained by selecting only the shared pairs or by averaging the results from different methods.

In this study we have also attempted to identify whether the difference in performance among covariation detection methods is due to phenomena of network connectivity among the covarying pairs. Several reports have stressed the importance of removing covariation due to "chaining" as a means to reduce false positive rates in the prediction of structural contacts [9-11]. This concept has been invoked again most recently in the introduction to the hpPCA method [19]. However, it is not clear that it is really the removal of chaining that produces better covariation maps. For example, in our testing of 9 protein families MI/mdMI based methods achieve close contacts recognition comparable to that of plmDCA and GREMLIN (**Figure S1**, Supporting Information) despite showing on average higher values of network transitivity.

Some clarity may be offered by considering what is the meaning of the partial correlation between variables (different positions in a sequence) in the context of the information that covariation detection methods are trying to extract from sequence data. In covariation studies, we typically start with some type of covariance matrix, and we try to guess its inverse, from which we can derive the partial correlations. This inverse can be related to the partial derivatives of a hidden optimization process that evolution carries out on a 'fit function' (which includes both functional and structural fitness), by changing one or more aa's at a time. From this point of view, the idea that transitivity depends on the existence of 'chains' of residues in which correlation is transferred from one residues to the next, so that distant residues appear correlated but in reality are not, loses appeal. The important question becomes instead: is a change in residue A by itself producing a change in the 'fit function F' (the partial derivative of F with respect to A) that goes in the same (or opposite) direction as a change in residue B by itself (the partial derivative of F with respect to B)? For two positions to appear correlated, it is not necessary to be part of a chain of contacts, as all that matters is their individual effect on the fit function. What is important in these cases of covariation is not the presence of a direct physical interaction, but the fact that residues exposed to like forces (e.g, the hydrophobic interior or the hydrophilic surface), will respond 'individually' (like a partial derivative) in a correlated (or anticorrelated/compensatory) way to the changes that affect the fit function.

On this basis, it appears that the reason why methods that derive partial correlation between the columns of a MSA (multi-dimensional MI, maximum entropy, sparse inverse covariance, pseudolikelyhood) provide a better recognition of close contacts is not because they remove chaining effects, but because they filter out the correlation between distant residues that originates from general fitness constraints [27] without the need for physical contacts. In contrast true chaining effects are expression of real physical perturbations that propagate inside proteins, and therefore are not removed by the derivation of partial correlation between variables.

## Methods

### MSAs.

MSAs for 9 protein families (the $F_1$ chaperone Atp11p [28], *p*-hydroxybenzoate hydroxylase (PHBH [29]), the catalytic subunit ArsA of the arsenic transporter [30], the $F_1$ chaperone Atp12p [28], phthalate dioxygenase reductase (PDR [31]), the arsenate reductase ArsC [32], KDO8P synthase, (KDO8PS [33]), CcrA (type 1 metallo-β lactamases [34]), and (*S*)-mandelate dehydrogenase (sMDH [35])) were calculated independently with T-Coffee [36] Muscle [37], and Mafft [38] and then merged with T-Coffee.

### Covariation detection methods.

A Matlab function (NMSA_to_mdMI) for the calculation of 3D_ and 4D_MI is available in the MSAvolve v2.0a Toolbox (download available from http://146.9.23.191/~gatti/coevolution/). The computation of 3D_MI and 4D_MI from equations (5,11) can be quite demanding as the number of combinations to be averaged rises very rapidly for large MSAs. To overcome this problem we have adopted a new MI algorithm, which was developed by Giangregorio Generoso to calculate the mutual information between two images, and was deposited at Matlab Central/File Exchange as the function 'MI_GG' (http://www.mathworks.com/matlabcentral/fileexchange/36538-very-fast-mutual-information-betweentwo-images/content/MI_GG.m). MI_GG is about 15 times faster than most versions of MI; while we provide in our Toolbox a version of the function optimized for the analysis of MSAs we refer to the original deposition for details of the algorithm.

In order to test the performance of PSICOV (which in the original implementation [14] has convergence problems in the GLASSO subroutine for the calculation of the sparse inverse when used with MSAs of <500 sequences), we recoded the algorithm as a Matlab function. The new function (NMSA_to_slPSICOV, also provided in our MSAvolve v2.0a Toolbox) does not show convergence problems due to the adoption of the QUIC algorithm [39] for the calculation of the sparse inverse.

GREMLIN, plmDCA, and hpPCA analyses were carried out with the original Matlab code downloaded from the authors' websites. For hpPCA the number *p* of patterns used in the calculation was calculated as $p = n\lambda 1 - n\lambda 2$, where $n\lambda 1$ is the total number of eigenvalues in the Pearson correlation matrix, and $n\lambda 2$ is the number of eigenvalues with value between 0.8 and 1.2 [19].

Merging of contact predictions obtained with different protein families and calculation of error margins (**Figure S1**) were carried out as described in [4].

### Distance dependent covariation signal.

A distance dependent covariation signal was used as a measure of how covariation scores change as a function of sequence distance. For each protein family, secondary structure assignments were obtained from the header of the PDB file of the reference X-ray structure. Then, given a secondary structure of length *n* comprising residues $r_1, r_2, r_3, r_4 ... r_{n-1}, r_n$, a *n*-1 by *n*-1 matrix of normalized covariation scores $C_{i,j}$ was constructed in which the 1st row contained the scores, $C_{1,1+1}$, $C_{1,1+2}$, $C_{1,1+3}$, $C_{1,1+4} ... C_{1,1+n-1}$, the 2nd row contained the scores $C_{2,2+1}, C_{2,2+2}, C_{2,2+3}, C_{2,2+4} ... C_{2,2+n-2}$, the 3rd row the scores $C_{3,3+1}, C_{3,3+2}, C_{3,3+3}, C_{3,3+4} ... C_{3,3+n-3}$, and so on until the *n*-1 row contained only the covariation score $C_{n-1,n}$ as its first element; all other elements of the matrix were set to 0. Average covariation scores for every sequence distance from 1 to *n*-1 were obtained by taking the average of the non-zero elements in each column of the matrix.

**Protein covariation graphs and network connectivity.**
The covariation matrix provided by each method corresponds to the weighted adjacency matrix of a protein covariation graph for each protein family. Given a threshold covariation score, a set number of top pairs from the covariation matrix is selected. These top pairs can be represented as a graph, where individual residues are nodes and edges exist between any two residues where the covariation score has exceeded the cutoff (i.e. the pair is in the top pairs). Two important factors contribute to the content of the covariation graph; i.) the number $n$ of pairs ii.) the minimum sequence distance $d$ between pairs. When selecting based on the number of pairs, $n$ is chosen as some fraction of the length, $L$ of the protein. When selecting a minimum sequence distance $d$, all covariation pairs for which the sequence distance is $< d$ are discarded from the graph. Covariation graphs were created for all protein families by varying either the graph size ($L$) or the minimum sequence distance ($d$). In practice, covariation matrices were first converted to the corresponding unweighted adjacency matrices, by setting all selected entries to 1 and all other entries (including the diagonals) to 0. Then, all connectivity analyses of the covariation graphs corresponding to these adjacency matrices were carried out with the Matlab Toolbox for Network Analysis developed by the Strategic Engineering Research Group (SERG) at MIT (http://strategic.mit.edu/).

## Authors' contributions
GWC and SHA carried out all the analyses. GWC, SHA, ERT, and DLG designed the analyses and wrote the manuscript.

## Acknowledgements
*Funding*: This research was supported by United States Public Health Service grants GM69840 to DLG, and GM48157 to SHA, by a Wayne State University Research Enhancement Program in Computational Biology grant to DLG, and by a Natural Sciences and Engineering Research Council of Canada (NSERC) grant to ERT.

**Figure Legends**

**Figure 1  - Correspondence between the distance map of (*S*)-mandelate dehydrogenase (sMDH) X-ray structure [PDB:1HUV] and the covariation maps for the MDH protein family obtained with different methods.**

Contact predictions by 2D_MI (light blue), 3D_MI (red), 4D_MI (black), PSICOV (blue), plmDCA (green), GREMLIN (cyan), and hpPCA (magenta) are shown as spots of size proportional to the covariation score. Gray regions represent the native distance map of the sMDH X-ray structure with a cutoff of 8 Å on the distance between the centroids of different residues.

**Figure 2  - Detection of close contacts by covariation maps.**

Covariation analysis was carried out for the MSAs of 9 protein families. For each family, in the panel on the left, each trace shows what percentage of all residue pairs separated by less than 8 Å in the reference X-ray structure is present in the top L covarying pairs identified by each method.  In the panel on the right, only pairs whose residues are separated by at least 20 intervening positions in sequence space are included in the analysis. True positives (covarying pairs corresponding to structural pairs < 8 Å apart) appear as upward displacements in the traces; false positives appear as horizontal segments in the traces. L ratio is the ratio between the number of sequences and the number of positions in each MSA.

**Figure 3  - Distance dependent covariation signal and secondary structure.**

For each secondary structure element (helices and strands), the average normalized covariation score was based on the sequence distance relative to an initial position. Curves connecting the average covariation scores at each sequence distance are drawn as cubic splines. Vertical yellow lines indicate the expected secondary structure period. The lower right two panels show the averaged scores for all 9 protein families.

**Figure 4  - Dependence of covariation scores on connectivity: MDH protein family**

**A. Dependence of covariation scores on path length. Top left.**  Mean length of the steps in each path connecting pairs within scoring threshold: the threshold is progressively moved to include a number of pairs equal to 3L. A vertical yellow line marks a number of pairs equal to L. **Top right.** Mean length of the path connecting pairs within scoring threshold. As more pairs are included in the analysis the probability of finding a shorter path increases and all traces converge to a smaller path length. **Bottom panels.** Correlation between the covariation score and the mean step length (**left panel**) or the total length (**righ panel**) of the path that connect the two members of each pairs through residues that belong to other pairs.
**B. Dependence of covariation scores on transitivity. Left panel .**  The size of the covariation graph is varied by including a progressively larger number of top scoring pairs at constant minimum sequence distance = 6. A vertical yellow line marks a number of pairs equal to L. **Right panel.** The minimum sequence distance is varied at constant graph size = L.

**Tables**

**Table 1 - Percentage of all pairs shared by different methods in the set of top L covariation scores: average of 9 protein families.**

|         | 2D_MI | 3D_MI | 4D_MI | PSICOV | plmDCA | GREMLIN | hpPCA |
|---------|-------|-------|-------|--------|--------|---------|-------|
| 2D_MI   | –     | 80.0  | 69.9  | 42.2   | 34.4   | 41.6    | 24.6  |
| 3D_MI   | 79.1  | –     | 87.9  | 46.4   | 36.9   | 45.0    | 27.5  |
| 4D_MI   | 67.7  | 86.3  | –     | 46.4   | 38.4   | 46.2    | 28.6  |
| PSICOV  | 38.3  | 41.6  | 42.1  | –      | 34.4   | 44.4    | 26.9  |
| plmDCA  | 29.0  | 31.5  | 32.4  | 32.6   | –      | 63.9    | 51.7  |
| GREMLIN | 36.4  | 40.0  | 40.2  | 41.9   | 61.2   | –       | 42.1  |
| hpPCA   | 20.2  | 22.6  | 23.8  | 23.2   | 46.2   | 37.6    | –     |

**Values above the diagonal.** All protein pairs. **Values below the diagonal.** Only pairs whose residues are separated by at least 20 intervening positions in sequence space.

**Figure 1.**

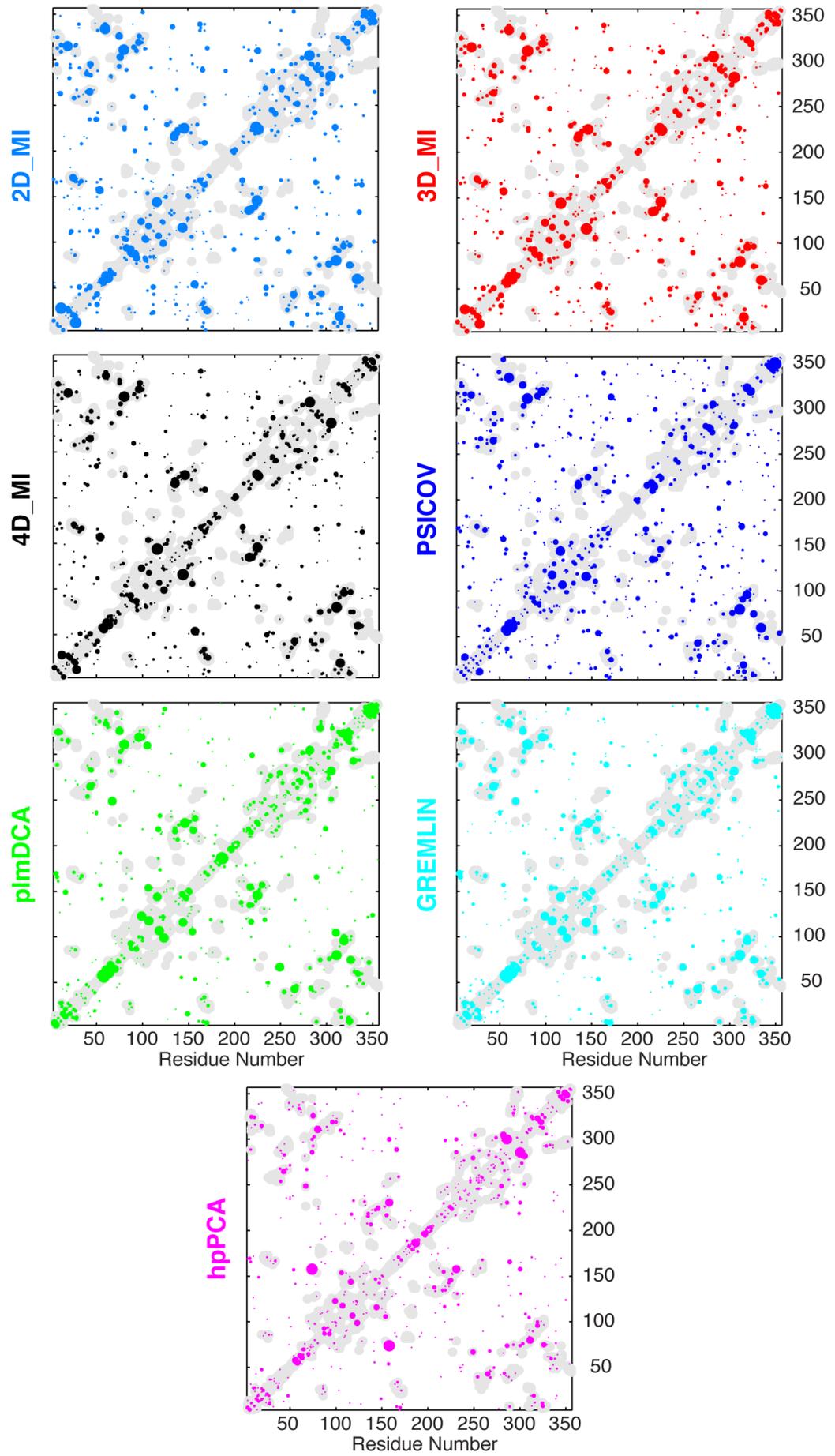

**Figure 2**

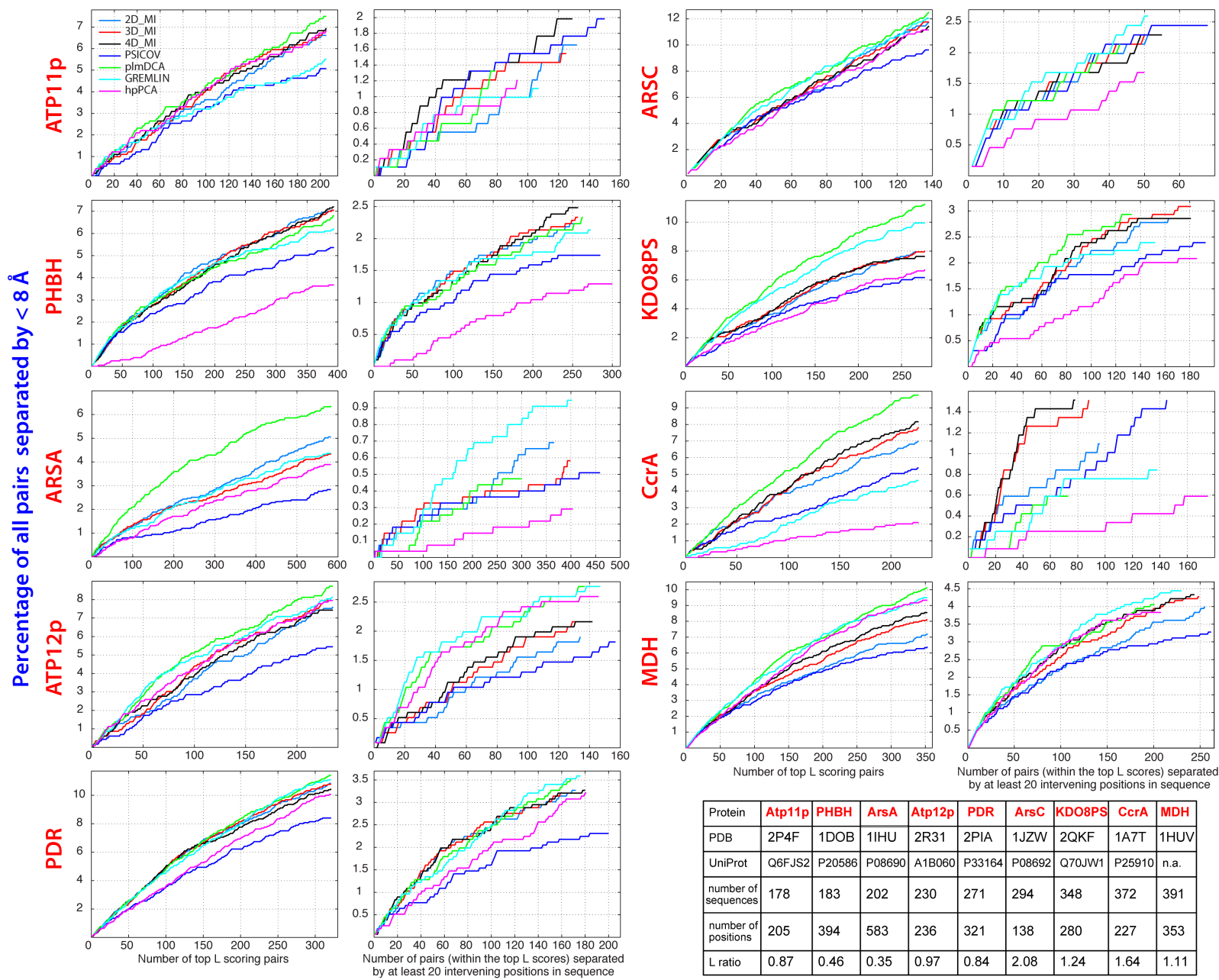

| Protein | Atp11p | PHBH | ArsA | Atp12p | PDR | ArsC | KDO8PS | CcrA | MDH |
|---------|--------|------|------|--------|-----|------|--------|------|-----|
| PDB | 2P4F | 1DOB | 1IHU | 2R31 | 2PIA | 1JZW | 2QKF | 1A7T | 1HUV |
| UniProt | Q6FJS2 | P20586 | P08690 | A1B060 | P33164 | P08692 | Q70JW1 | P25910 | n.a. |
| number of sequences | 178 | 183 | 202 | 230 | 271 | 294 | 348 | 372 | 391 |
| number of positions | 205 | 394 | 583 | 236 | 321 | 138 | 280 | 227 | 353 |
| L ratio | 0.87 | 0.46 | 0.35 | 0.97 | 0.84 | 2.08 | 1.24 | 1.64 | 1.11 |

**Figure 3**

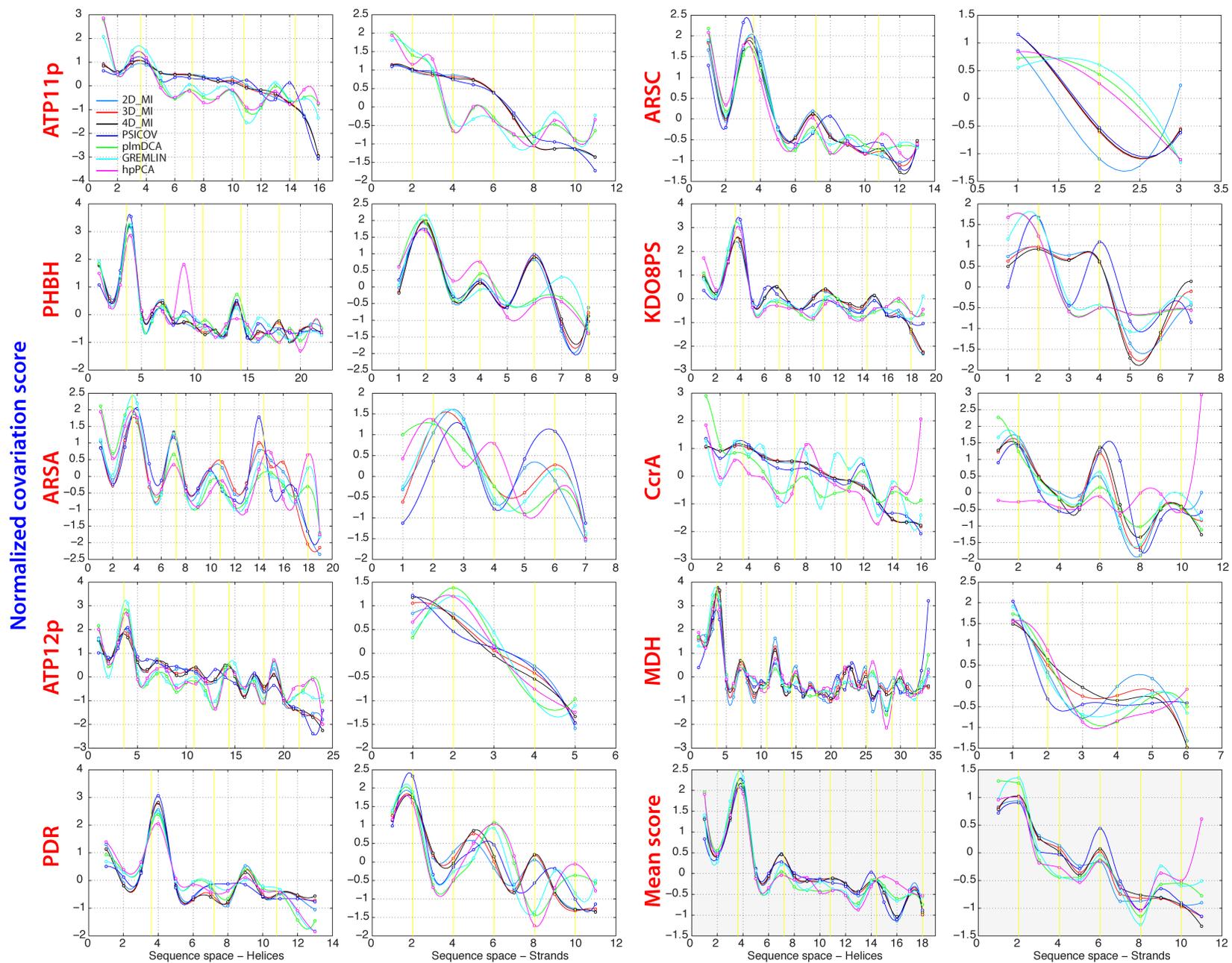

**Figure 4**

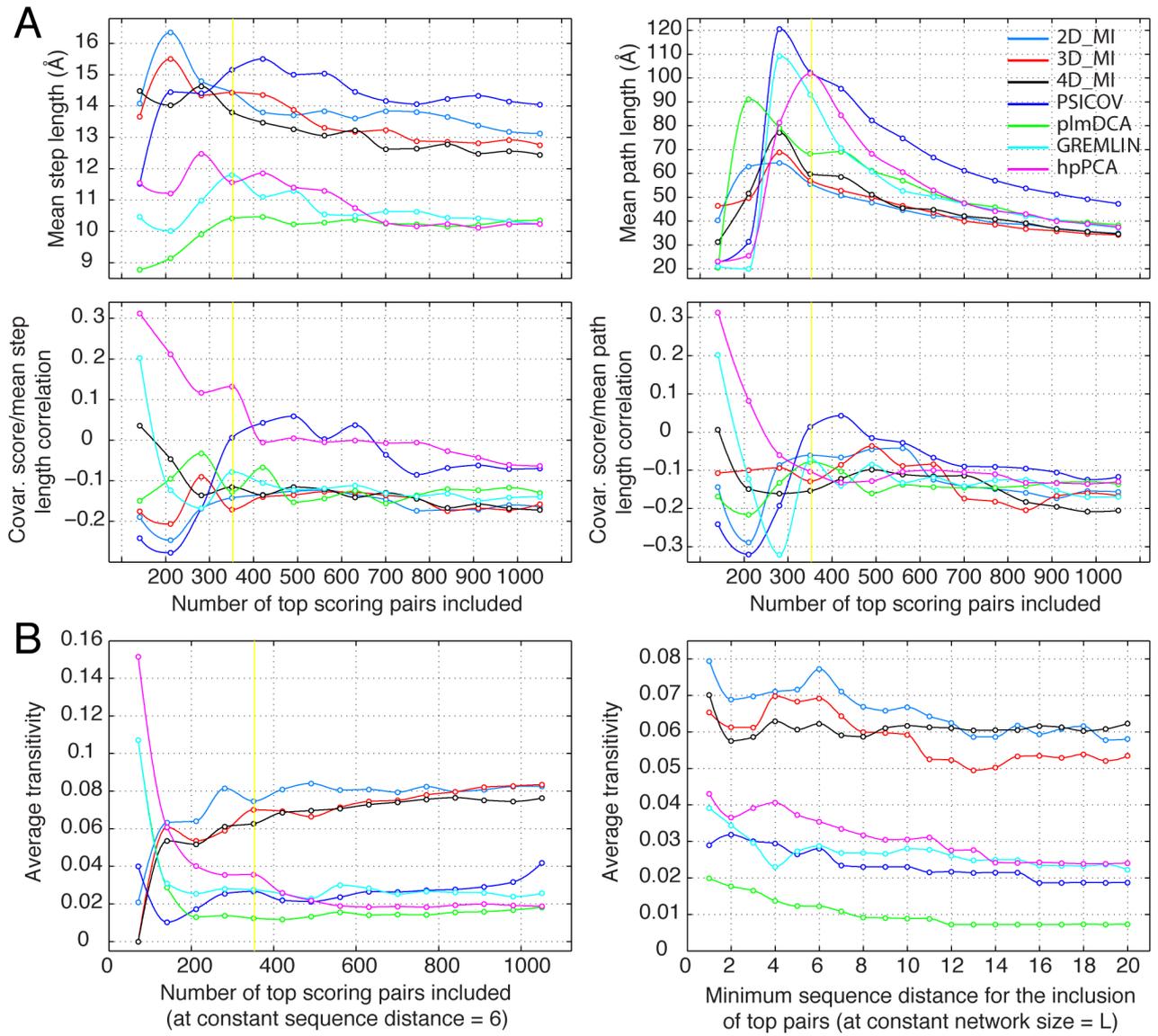

# Supporting Information

**Figure S1** - Detection of close contacts by covariation maps: averaged results for all protein families.

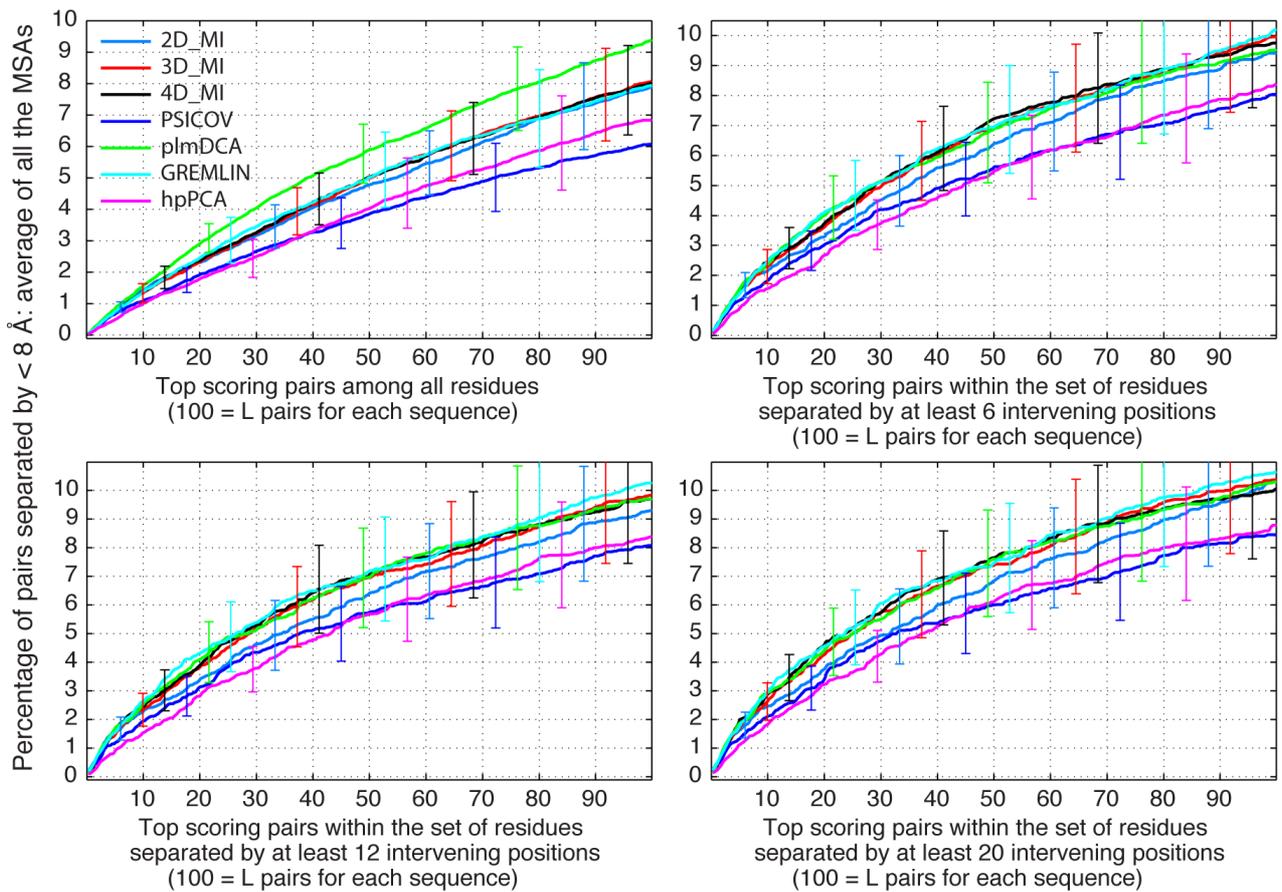

Averaged results for the set of 9 protein families. Each panel shows what percentage of the residue pairs separated by less than 8 Å in the X-ray structure of each protein is present on average in the top covarying pairs identified by each method. The abscissa scale is normalized such that 100 corresponds to a number of pairs equal to the number of residues in each sequence. **A.** All protein pairs. **B,C,D.** Only pairs whose residues are separated by at least 6, 12, 20 intervening positions in sequence space.

**Figure S2 - Dependence of covariation scores on path length connectivity: average of all protein families.**

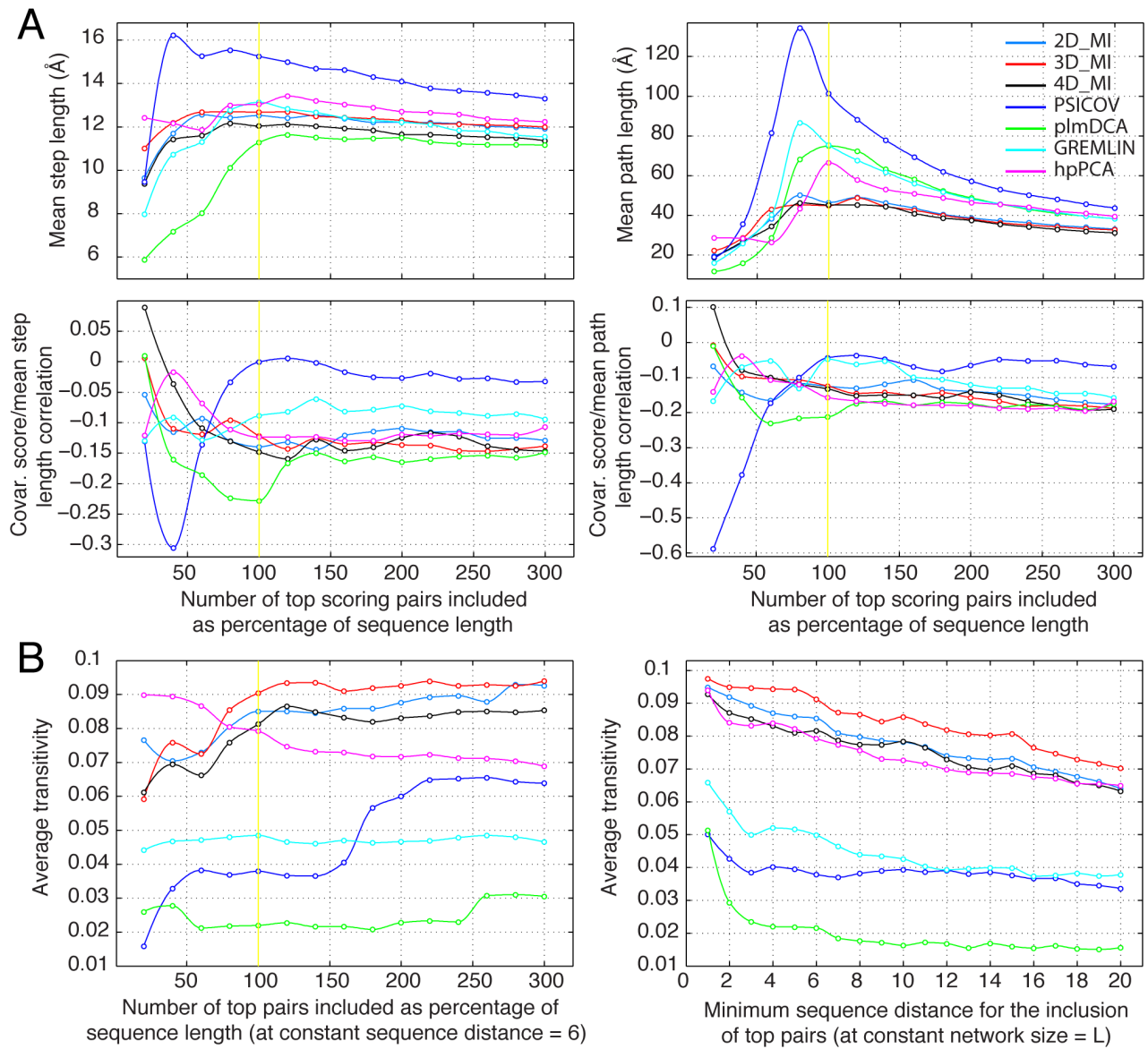

**A. Dependence of covariation scores on path length. Top left.** Mean length of the steps in each path connecting pairs within scoring threshold. The threshold is progressively moved to include a number of pairs equal to 3L. A vertical yellow line marks a number of pairs equal to L. **Top right.** Mean length of the path connecting pairs within scoring threshold. As more pairs are included in the analysis the probability of finding a shorter path increases and all traces converge to a smaller path length. **Bottom panels.** Correlation between the covariation score and the mean step length (**left panel**) or the total length (**righ panel**) of the path that connect the two members of each pairs through residues that belong to other pairs.

**B. Dependence of covariation scores on transitivity. Left panel.** The size of the covariation graph is varied by including a progressively larger number of top scoring pairs at constant minimum sequence distance = 6. A vertical yellow line marks a number of pairs equal to L. **Right panel.** The minimum sequence distance is varied at constant graph size = L.

**Text S1 - Derivation of 4-dimensional MI.**

Mutual information $I_{X3,X4}(X1; X2)$ between X1 and X2, when the effect of two additional variables X3 and X4 on the transmission between them is removed, is obtained as:

$$I_{X3,X4}(X1; X2) = \sum p_{x3,x4} \, I(X1; X2 \mid X3 = x3, X4 = x4) = I(X1; X2 \mid X3, X4) \qquad (6)$$

By the 'chain property' of multivariate MI we derive:

$$I(X1; X2 \mid X3, X4) = I(X1; X2 \mid X4) - I(X1; X2; X3 \mid X4)$$
$$= I(X1,X2) - I(X1;X2;X4) - I(X1;X2;X3) + I(X1;X2;X3;X4) \qquad (7)$$

where the 'interaction information' between the four variables is:

$$I(X1;X3;X2;X4) = [H(X1) + H(X2) + H(X3) + H(X4)] - [H(X1,X2) + H(X1,X3) + H(X1,X4) + H(X2, X3) + H(X2,X4) + H(X3,X4)] + [H(X1,X2,X3) + H(X1,X2,X4) + H(X1,X3,X4) + H(X2,X3,X4)] - H(X1,X2,X3,X4) \qquad (8)$$

Averaging over all values of X3 and X4 (two additional columns of the MSA), and recalling that all the values taken by X3 and X4 are the same with respect to X1 and X2 in a MSA, we finally obtain:

$$<I_{X3,X4}(X1;X2)>_{X3,X4 \neq X1,X2} = I(X1;X2) - 2<I(X1;X2;X3)>_{X3 \neq X1,X2} +$$
$$<I(X1;X2;X3;X4)>_{X3,X4 \neq X1,X2} \qquad (9)$$

Expanding (7) leads to a direct expression of $I_{X3,X4}(X1;X2)$ in terms of the entropies of the individual variables:

$$I_{X3,X4}(X1;X2) = H(X1) + H(X2) - H(X1,X2)$$
$$- [H(X1) + H(X2) + H(X4) - H(X1, X2) - H(X1, X4)$$
$$- H(X2, X4) + H(X1, X2, X4)]$$
$$- [H(X1) + H(X2) + H(X3) - H(X1, X2) - H(X1, X3)$$
$$- H(X2, X3) + H(X1, X2, X3)]$$
$$+ [H(X1) + H(X2) + H(X3) + H(X4)] - [H(X1,X2) + H(X1,X3) + H(X1,X4)$$
$$+ H(X2, X3) + H(X2,X4) + H(X3,X4)] + [H(X1,X2,X3) + H(X1,X2,X4)$$
$$+ H(X1,X3,X4) + H(X2,X3,X4)] - H(X1,X2,X3,X4)$$

$$= \cancel{H(X1)} + \cancel{H(X2)} - \cancel{H(X1,X2)}$$
$$- \cancel{H(X1)} - \cancel{H(X2)} - H(X4) + \cancel{H(X1, X2)} + H(X1, X4)$$
$$+ H(X2, X4) - \cancel{H(X1, X2, X4)}$$
$$- \cancel{H(X1)} - \cancel{H(X2)} - H(X3) + \cancel{H(X1, X2)} + H(X1, X3)$$
$$+ H(X2, X3) - \cancel{H(X1, X2, X3)}$$
$$+ \cancel{H(X1)} + \cancel{H(X2)} + H(X3) + H(X4) - \cancel{H(X1,X2)} - H(X1,X3) - H(X1,X4)$$
$$- H(X2, X3) - H(X2,X4) - H(X3,X4) + \cancel{H(X1,X2,X3)} + \cancel{H(X1,X2,X4)}$$
$$+ H(X1,X3,X4) + H(X2,X3,X4) - H(X1,X2,X3,X4)$$

which simplifies to:

$$I_{X3,X4}(X1;X2) = - H(X3,X4) + H(X1,X3,X4) + H(X2,X3,X4) - H(X1,X2,X3,X4) \qquad (10)$$